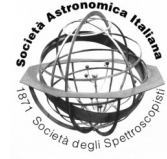

# Infrared emission from gravitational wave sources with THESEUS/IRT

S. Piranomonte[1]

Istituto Nazionale di Astrofisica – Osservatorio Astronomico di Roma, Via Frascati 33, I-00078 Monte Porzio Catone, Rome, Italy  e-mail: `silvia.piranomonte@oa-roma.inaf.it`

**Abstract.** With the discovery of the electromagnetic counterpart of the gravitational wave source GW170817 the multi-messenger era is started. The identification of an electromagnetic counterpart is crucial to understand the nature of the detected gravitational wave sources and to maximize the scientific return of their detections. The role of the instrument THESEUS/IRT will be crucial in this field, in particular in localizing afterglows of gamma-ray bursts within few minutes from the trigger and in identifying optical/NIR isotropic emissions such as kilonovae.

## 1. Introduction

The gravitational wave (GW) emitter candidates detectable by terrestrial interferometers in the frequency range of 10-10000 hz are the mergers of binary systems hosting two neutron stars (NS), a neutron star and a black hole (NSBH) or two black holes (BBH), i.e. the core collapse of massive stars with large degree of asymmetry and fast rotating asymmetric isolated NSs. The merger of two NSs or a NS with a BH and the core collapse of massive stars are expected to produce short and long Gamma-Ray Bursts (GRBs) respectively, while soft gamma-ray repeaters and/or magnetars are expected to be associated with the isolated NS. Since now, two observing runs (O1 and O2) have been carried out by three interferometers LIGOs and Virgo. Virgo joined the two LIGOs during the last month of O2 (August 2017). During these runs 7 events were detected: 6 secure detection, GW150914 (Abbott et al. 2016a), GW151226 (Abbott et al. 2016b), GW170104 (Abbott et al. 2017a), GW170814 (Abbott et al. 2017a), GW170608 (Abbott et al. 2017c) and GW170817 (Abbott et al. 2017d) and one low-significance candidate LVT 151012 (Abbott et al. 2016c). The analysis of the first five GW signals revealed that all these events were originated from the inspiral and merger of binary black hole systems while the last one, GW170817, was associated for the first time to the binary merger of two neutron stars. The spectacular detection of the first electromagnetic (EM) counterpart of the GW event originated by the coalescence of a double neutron star system (GW 170817) has demonstrated that, thanks to a huge observational effort, it is possible to observe the electromagnetic emission of a GW source. In this particular case, the astronomical community involved in the search of the electromagnetic counterparts of GW sources have proved, for the first time, that binary neutron star merger is associated with at least a short gamma-ray burst (SGRB) with faint gamma-ray pulse and late-time X-ray emission. Some studies explained this behaviour with an afterglow emission which could be due to a structured jet observed off-axis (e.g. Troja et al., 2017; Alexander et al., 2017; Margutti et al., 2017; Haggard et al., 2017; Hallinan et al., 2017; Lazzati et al., 2017) or to the deceleration of an isotropic middle relativistic outflow (Mooley et al. 2017; Salafia et al. 2017). This is suggesting that we are witnessing an extension of the SGRB population also to different, fainter and nearer events (i.e. ≤ 100 Mpc). Moreover, several multiwavelenght observations from radio to NIR of this event provided the first compelling observational evidence for the existence of kilonovae, transient sources powered by radioactive decay of heavy elements resulting from the r-process nucleosynthesis of ejected neutron star matter (i.e. Pian, D'Avanzo et al. 2017; Abbott et al. 2017b; Tanvir et al. 2017; Nicholl et al. 2017; Smartt et al. 2017; Tanaka et al. 2017; Chornock et al. 2017).

With this first detection, it is clear that the EM radiation associated with NS-NS is emitted at all frequency bands and that different emissions peak at different times from the merger, from several seconds before and, most probably, to few years after, therefore it is important to have a network of multi-wavelenght observatories which can cover huge regions of the sky and repeat observations over different timescales.

## 2. The merger of two neutron stars and their infrared emission

The merger of two neutron stars or a neutron star and a black hole have several astrophysical implications: they seems to be the engine behind the short gamma-ray bursts, they produce the heaviest elements in the universe, the radioactive decay of heavy elements causes an electromagnetic transient ("kilonova") that accompanies the expected gravitational wave signal.

During the NS-NS or NS-BH mergers a certain amount of ejected mass is expected to become unbound. There are two main different channels of mass ejection in a BNS/NS-BH merging: the dynamical ejecta or the so-called "red kilonova" (unbound by hydrodynamic interaction and gravitational torques) and the disk wind outflows (blue kilonova) which can have different physical origins. They could be driven off by neutrino absorption, by magnetically launched winds or by accretion disk matter that becomes unbound by viscous and nuclear heating. While the dynamical ejecta are expected to be highly neutronized, with an electron fraction $Ye = np/(np + nn) \leq 0.1$ (with $np$ and $nn$ the proton and neutron number density), the situation is different for the wind outflows whose neutronization degree depends on the neutrino irradiation from the system (Kasen et al. 2015). This in turn depends on the nature of the central compact remnant of the merger. If a fast rotating BH is formed, the accretion disk will irradiate more neutrinos than in the slow rotating case, due to the smaller size of the ISCO (innermost stable circular orbit) that implies hotter inner disk. If otherwise a HMNS (high-mass neutron star) is formed, the irradiation will be even stronger and will rise with the star lifetime. Higher neutrino irradiation will imply higher Ye, which in turns quenches the lanthanides nucleosynthesis, lowers the opacity and thus produces a bluer emission (blue kilonova). In any case the faster and earlier launched dynamical ejecta are expected to act as a lanthanides, high opacity, curtain that will almost suppress this blue transient (Kasen et al. 2015). The blue kilonova should peaks one day after the merger while we expect to see the red one at ∼ days - 1 week after the merger. Both channels eject extremely neutron-rich matter. These channels are sketched in Fig. 1. Several authors tried to study the radioactively powered transients produced by accretion disc winds following a compact object merger in order to generate synthetic light curves and spectra. Kasen et al 2015 showed that resulting kilonova transients generally produce both optical and in-frared emission and that the kilonova light curves thus typically has two distinct components (see their Figure 4), while thermal spectra evolve from optical to NIR due to different opacities predicted in the disc outflows and in the middle-relativistic ejecta (see Barnes et al. 2013).

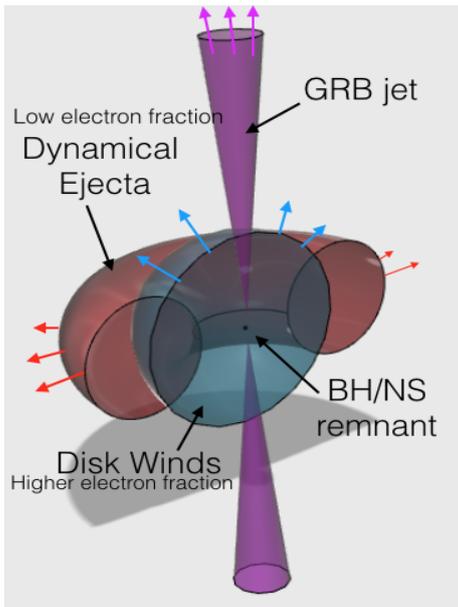

**Fig. 1.** A cartoon of the kilonova eject material via different channels: the "dynamic ejecta" (unbound by hydrodynamic interaction and gravitational torques) and various types of "winds"; credits: S. Ascenzi.

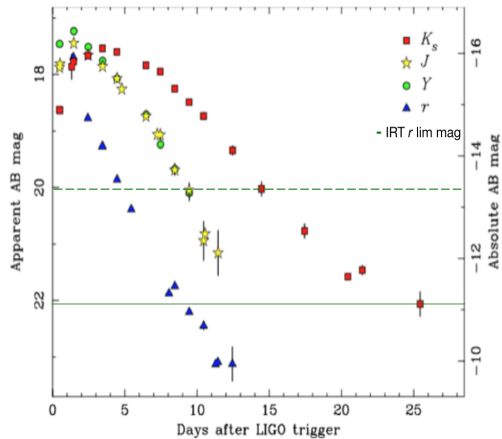

**Fig. 2.** Light curves of AT2017gfo in the r-,Y-,J-, and Ks- bands together with the limiting $r$ magnitudes achievable with THESEUS/IRT (green lines). Dashed line for the spectroscopy, solid line for the imaging. Adapted from Tanvir et al. 2017.

### 2.1. The discovery of the kilonova GW170817

The 17th of August 2017 the first NS-NS GW event was detected by aLIGO/Virgo (GW 170817) and it was associated to the weak SGRB 170817A detected by the Fermi and Integral satellites (Abbott et al 2017e; Goldstein et al., 2017; Savchenko et al., 2017). The proximity of the event (∼ 40 Mpc) and the relative accuracy of the localization (∼ 33 deg$^2$, thanks to the joint LIGO and Virgo operations) led to a rapid ($\delta t \leq 11h$) identification of a relatively bright (i ∼ 17.3 mag) optical counterpart, AT2017gfo (also known as SSS17a; Coulter et al., 2017) in the galaxy NGC 4993. Following the detection of this source, several imaging and spectroscopic follow-up campaign at optical and NIR wavelengths were performed (see for example the observed NIR light curve from Tanvir et al. 2017 in Fig.2 and the spectral evolution of AT2017gfo from Pian, D'Avanzo et al. 2017 in Fig. 3).

Pian, D'Avanzo et al. 2017 showed that the analysis and modelling of the spectral characteristics of this source, together with their evolution with time, result instead in a good match with the expectations for kilonovae, providing the first compelling observational evidence for the existence of such elusive transient sources. Comparison with spectral models suggests the merger ejected 0.03-0.05 solar masses, including high-opacity lanthanides. The single (by now) case of GW 170817, besides representing an historical result, clearly demonstrates the vital importance of time-resolved optical/NIR imaging and spectroscopic follow-up of electromagnetic candidate counterparts of GW triggers.

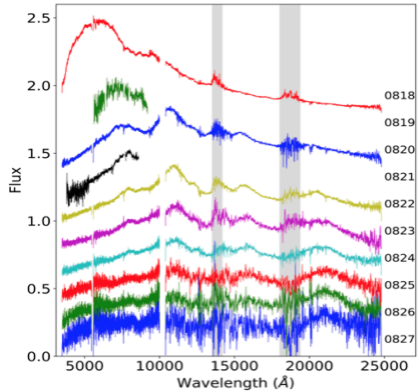

**Fig. 3.** Time evolution of the AT2017gfo spectra. VLT X-shooter and FORS2 spectra are shown, the flux normalization is arbitrary (adapted from Pian, D'Avanzo et al. 2017).

## 3. Core collapse of massive stars: Long GRBs, Low Luminosity GRBs and Supernovae

The collapse of massive stars are expected to emit GWs since a certain degree of asymmetry in the explosion is present. The collapsar scenario invoked for long GRBs (e.g. Woosley 1993, Paczynski 1998) requires a rapidly rotating stellar core, so that the disk is centrifugally supported and able to supply the jet. This rapid rotation may lead to non-axisymmetric instabilities, such as the fragmentation of the collapsing core or the development of clumps in the accretion disk. The association of nearby long GRBs with the core collapse SNe (CCSNe, e.g. Woosley and Bloom, 2006; Galama et al., 1998; Stanek et al., 2003) implies that any closeby long GRB, should be associated with a detectable GW emission and thus offers a very interesting potential synergy between EM facilities (from gamma to radio band) and GW detectors.

## 4. The role of the instrument IRT

The InfraRed Telescope on-board THESEUS (IRT, 0.7-1.8 $\mu m$) is a 0.7 m class IR fast response telescope with 10×10 arcmin FOV with both imaging and spectroscopy capabilities. Regarding the GW sources, THESEUS will trigger and localize transient sources within the uncertain GW and/or neutrino error boxes with its instruments XGIS and/or with SXI. After a SXI/XGIS trigger, if an optical counterpart is present, the sky localization of the source can be refined down to few arcseconds with IRT. If bright enough, spectroscopic observations could be performed onboard, thus providing redshift estimates and information on chemical composition of circumburst medium. If the optical/NIR counterpart is not bright enough for performing on-board spectroscopy, precise sky coordinates will be circulated to ground based telescopes in order to perform follow-up observations. In particular, for collimated EM emission from short GRBs, IRT could point the SXI localized afterglow within few minutes from the trigger and for the optical/NIR isotropic emissions IRT will be optimal to identify the kilonova and to disentangle the different components due to its photometric and spectroscopic capabilities (see Fig 7 and 8 of Stratta et al. 2017).

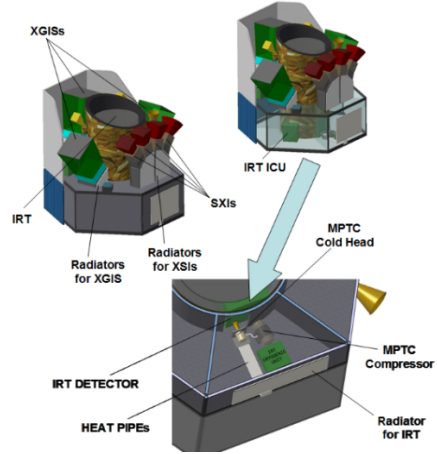

**Fig. 4.** A sketch of the THESEUS satellite and the accommodation of the IRT instrument.

Furthermore, for LGRBs, Low Luminosity GRBs and Supernovae the off-axis X-ray afterglow detections (orphan afterglows) can potentially increase the simultaneous GW+EM detection rate by a factor which strongly depends on the jet opening angle and the observer view-ing angle. THESEUS with IRT may also observe the appearance of a NIR orphan afterglow few days after the reception of a GW signal due to a collapsing massive star. In addition, the possible large number of low luminosity GRBs (LLGRBs) in the nearby Universe, expected to be up to 1000 times more numerous than long GRBs, will provide clear signatures in the GW detectors because of their much smaller distances with respect to long GRBs.

## 5. Conclusions

In the 2020s we expect several synergies at different wavelengths such as:

- the space-based telescopes James Webb Space Telescope (JWST), ATHENA and WFIRST;
- the ground-based telescope with large FOV like zPTF and LSST which will be able to select the GW candidates in order to follow-up them afterwards;
- the large multi-wavelengths telescopes such as the 30-m class telescopes GMT, TMT and E-ELT, which will all follow-up the optical/NIR counterparts like kilonovae;
- the Square Kilometer Array (SKA) in the radio, which is well suited to detect, for example, the late-time (∼weeks) signals produced by the interaction of the ejected matter with the interstellar medium.

THESEUS/IRT will be perfectly integrated in this search, thanks to its photometric and spectroscopic capabilities and its spectral range coverage, allowing us to improve the study of such (until now) uncertain emission mechanisms. With IRT we will acquire not only the light curves of EM counterparts but also their spectra, thus having the opportunity to disentangle the different components associated to some of those events (e.g. kilonova emissions like disk wind + dynamical ejecta).

Third generation GW detectors such as the Einstein Telescope should arrive to detect at luminosity distances of order 1 Gpc (e.g. Davies et al. 2002). The first joined GW/GRB/SN observations, possibly combined also with neutrino detections, will prove crucial to unravel the nature of these sources and their explosion mechanism. THESEUS will likely play a very important role in this investigation. For any further details about THESEUS detection capabilities in this field, see the multi-messenger astrophysics THESEUS paper (Stratta et al. 2017).